\documentclass{article}

\begin{document}

{\bf DISCRETE SYMMETRY IN THE EPRL MODEL AND NEUTRINO PHYSICS}

\bigskip

by: Louis Crane, Mathematics department, KSU.

\bigskip

{\bf ABSTRACT:} {\it In \cite{C1}, we proposed a new 
interpretation of the EPRL quantization 
of the BC model for quantum general relativity using a monoidal functor
we call the time functor.

In this preliminary draft we apply the theory of modules over monoidal functors
\cite{Y1} to the time functor, to propose an extension of the EPRL 
model which would
include the standard model. This is motivated by recent advances in
neutrino Physics.}

\bigskip

{\bf Introduction. The algebraic structure of the EPRL model, and new
  approaches to constructing realistic matter fields. }

\bigskip
This paper begins an attempt to construct a
grand unified theory, using categorical quantum geometry. The idea is
to extend the EPRL model for quantum general relativity using module
categories and functors.

In \cite{BC} a new approach to the quantum theory of gravity was
begun. The theory modelled spacetime on a simplicial complex, rather
than a smooth manifold based on a point set continuum. Structural similarities
between the algebra of tensor categories and the combinatorial
topology of simplicial complexes were used to put a quantum geometry
on the complex. This model had a number of successes, but did not have
a good classical geometrical limit or Hilbert space on a space slice.

Aside from the well known renormalization problems of general
relativity and the Planck length, this was motivated in part by the
frustrating outcomes of attempts to get realistic matter fields from
classical continuum geometries.

The question of extending the BC model to include realistic matter fields did
not make much headway.

In \cite{EPR},  \cite{LS} \cite{EP} and \cite{B}, a new version of the
BC model has been developed. The problems relating to the Hilbert
space and the geometrical interpretation have been resolved.

In \cite{C1}, the author described the mathematical form of the
new model. The weak constraints were interpreted as given by a functor
from Rep(SO(3)) to  Rep(SO(4)) in the Euclidean signature, and to
Rep(SO(3,1)) in the Lorentzian signature, which we called the {\bf
  time functor}. The physical Hilbert space
is just the preimage of the time functor. 

An interpretation of the model was proposed in which the universe was
in a topological state, which meant that the domain category of $F_t$
was also used to construct a three dimensional topological field
theory. This suggests that the new version of quantum gravity is
related to the categorical constructions of TQFTs in various
dimensions \cite{CY}, \cite{CF}. 

In this paper, we consider extensions of the EPRL model which seem to
be promising ways to reproduce the standard model.

Our motivation for this is a suggestive coincidence between the theory
of module functors over the time functor and recent developments in
neutrino Physics. Recent attempts to model neutrino oscillations
together with quark mixing matrices yield structures which also appear
in the mathematical analysis of modules over $F_t$. In particular, in
the work of Ma \cite{Ma} the discrete group of symmetries of the
tetrahedron and the exceptional Lie algebra $E^6$ \cite{Ma1} both appear in
the mathematical foundation of a unified model which would explain the
three generations with neutrino oscillations included. These two
structures are linked in the 
correspondence between discrete subgroups of the rotation group and
the simply laced Lie groups due to Mckay \cite{Mac}, which is closely related
to the theory of module categories over the category of
representations of Rep (SO(3)) and also of its quantum group deformation.

We can formulate this more suggestively by thinking of the data in a state sum 
as a substitution for the
geometry and topology of the spacetime manifold. To include Yang-Mills fields and
fermionic matter, we must find a categorical equivalent for bundles
over the manifold. Bundles appear in more algebraic approaches to
topology as modules over the base space; for example, the space of cross sections
of a bundle forms a module over the function space of the manifold. 
Thus modules over the two categories plus functor out of
which the EPRL model can be formed are a reasonable candidate for
matter fields.  

A module functor over the time functor would include module categories
over its domain and range, and a functor analogous to the time functor
between them. This is explained below. This would allow us to
construct a state sum model analogous to the EPRL model from it. The
image of the functors down from the module categories to the original
pair would define a gravitational sector of the new model. 

In the rest of this paper we give expositions of the mathematical
structures we are considering, which we hope will be accessible to the
quantum gravity community, and explore several possible ways to use them to
produce a grand unified theory.

\bigskip

{\bf The Time Functor}

\bigskip

The point of departure for the new model was the discovery of an
elegant way to impose the simplicity constraints of the BC model
weakly rather than strongly \cite{EPR}. The EPR physical Hilbert space
is composed of vectors in Rep(SO(3)) so the Hilbert space on a
3-manifold comes from the preimage of the relativistic representations
labelling it under a functor. 

The EPRL model has both a Euclidean and a Lorentzian signature
version. Each can be described as a new type of functor, but they are
quite different. In the
Euclidean case, the functor

\bigskip

$F_t: Rep SO(3) \rightarrow Rep SO(4) $

\bigskip

can be defined by

\bigskip

$F_t(R_{2i})=R_{i,i}$

\bigskip

(where i is a half integer) and expanding linearly on direct sums. It follows fron the elementary
theory of spins that there exists a natural map

\bigskip

$M_{a,b}: F(a \otimes b) \rightarrow F(a) \otimes F(b)$.

\bigskip

M is an injection. Since we are in a category of Hilbert spaces, we
can equally well define the projection $M^*$ which goes backward. In
the terminology of category theory, we can equally well say the functor
is lax, using $M^*$.

The functor $F_t$ may be only weakly colax, however, by which we mean that
the associators in SO(3) and SO(4) do not agree, so a correction
factor must be included, connecting them in the coherence pentagon. We
are still investigating if the correction can be made in the tensor
operators of the category, so that the amended category is strictly colax.

In the Lorentzian signature case, the EPRL model proceeds by
reducing the irreducible
representations of the Lorentz group which are the building blocks of
the theory into sums of representations of a suitable copy of SU(2)
and selecting only the lowest spin representation which appears. 

Since the Lorentz group is noncompact, the decompositions of tensor
products are direct integrals. To make the following discussion
mathematically rigorous, we will need a richer categorical description
of Rep(SO(3,1)) in which objects are direct integrals of simple
representations under general measures and morphisms include singular
maps connecting $L^2$ and distributional objects like
the theory of rigged Hilbert spaces, a
further development of the theory of measured categories. We shall
proceed somewhat formally in what follows.

Since
an important aspect of the EPRL model is that it relates the
Hamiltonian picture of loop quantum gravity, in which states are
described by spin networks in space, with relativistic spin networks
in spacetime; we want to think of it as a rule which assigns `` spacetime''
representations of SL(2,C) to ``space'' representations of
SO(3,R). The proper mathematical expression of this is a functor 
\bigskip

$ F_{\gamma } : REP(SO(3,R)) \rightarrow REP (SL(2,C)); $ 

\bigskip

Defined by:

\bigskip

$F_{\gamma}(R_k)= R(k, \gamma k)$.
\bigskip

 which depends on the Immirzi parameter $\gamma $. The case $\gamma =0 $
is a degenerate case in which much of what follows is incorrect. 

This functor assigns to each irreducible representation $R_k$ of
SO(3,R) the irreducible $R(k, \gamma k)$ of SL(2,C). Since the only
morphisms between irreducibles in either category are multiples of the
identity, the action of the functor on morphisms is immediate.

Considerably more subtle are the tensor and ``renormalizability''
properties of this functor, which are generalizations of the very
special facts which make the Lorentzian EPRL model finite and
physically interesting.

The action of the time functor on direct sums is straightforward. The
behavior on tensor products is more subtle. The image under $F_\gamma
$ of the tensor product of two objects injects into the tensor product
of the two images.
\bigskip

(1)  $ F_{\gamma }( X \bigotimes Y) \rightarrow F_{\gamma } X \bigotimes
F_{\gamma } Y$.

\bigskip

As we explained above, equation 1 is expressed by mathematicians by
saying that the functor is  (possibly weakly, as in the Euclidean case) colax.
 
However, the injection is improper, in the sense that a dirac delta
function is an improper object of $L^2(R)$. This is because
REP(SL(2,C)) contains representations labelled by a continuous
parameter, 
and the tensor product of two
representations is a direct integral in the sense of Mackey or
Gelfand.  

In order to make the construction of the time functor rigorous, we
need a richer description of the category Rep(SO(3,1)) which would
include distributional morphisms. This will require a further
development of the theory of measured categories \cite{MC}. We
therefore proceed somewhat formally in the following.

One might also like to take the formal projection dual to the
injection in (1), but it is also too singular, and could be made rigorous
only in the sense of a proper value. Further technical foundational
work will be needed to formulate this.

$F_{ \gamma}$ connects the discrete spectrum of areas in loop quantum
gravity or 3d TQFT with the continuum of representations in spacetime
models. The geometrical data on tetrahedra in the EPRL model
correspond to intertwiners in SO(3,R), lifted by the time functor, and
not to general SL(2,C) intertwiners.

Since the image category consists of infinite dimensional
representations, there is no hope of the trace in the domain category
going over via the functor to the range. Thus a naive evaluation of
closed diagrams, such as the 15J symbols in the model is impossible. However, there is a
renormalizable trace which is well defined on an ample set of
diagrams in the domain category.

This renormalizable trace is just the multiple integral over SL(2,C) [3,4], 
which played a crucial role in the finiteness of the BC model, and
goes over to the new model. The renormalization just consists in
dropping one of the integrations (it doesnt matter which). The
finiteness of the resulting integral expression is the key to the
success of both models.

The ample set of diagrams on which the renormalized trace is finite
includes the free graph on 4 or 5 vertices, which represent the
tetrahedron and 4-simplex, and includes any diagram obtained by adding
edges to any diagram already in the ample set \cite{JJ}.

We can provisionally define a {\bf claren functor}
(co-lax, amply renormalizable) between
two tensor categories as one with an inclusion as in equation 1 for
any pair of objects X,Y in the first category, and a regularizable
trace for the images under the functor of an ample family of diagrams in the 
second category as defined above.

\bigskip

{\bf Neutrino Oscillations and their implications for unified models}

\bigskip

Recent research has shown that the three different types of neutrinos
are coupled. Surprisingly, their couplings involve large angles,
unlike the mixing matrix for the quarks which involves smaller angles.

In attempting to explain this, it has been suggested \cite{Ma} that the
elementary particles, in addition to living in representations of Lie
groups which explain their Yang-Mills charges, also live in
representations of finite groups. In order to produce three 
weekly mixed pairs of quarks with widely differing masses and three
strongly mixing neutrinos, the group needs to have three one
dimensional representations 1, 1', and 1'', and a three dimensional
representation 3.  A search among finite groups yielded $A_4$, the
symmetry group of the tetrahedron, as the natural candidate. Coupling
matrices to Higgs sectors which are invariant under this group yield
the tribimaximal mixing matrix, which has emerged from a long series
of delicate experiments as a strong candidate for the neutrino sector.
The discrete group symmetry is an advance in that the three
generations of fermions are no longer just put in by hand.

Another line of research \cite{Ma1} has been in looking for grand unified groups
for theories which include neutrino masses and neutrino mixing. It is
difficult to get the quark and neutrino sectors to break in such
different ways without two separate $U_1$'s which suggests a rank 6
group \cite{Ma1}. $E^6$ seems to be the most attractive candidate.  

So putting this together, a combined symmetry under the discrete group
$A_4$ and the Lie Group $E^6$ seems to be a likely setting for a grand
unified theory which includes the latest neutrino Physics.

Can we find an explanation of the combination $A_4 \times E^6$ in the
context of our proposal?

\bigskip

{\bf Module categories and functors, and discrete subgroups}

\bigskip

In \cite{CF} it was observed that the direct sum and tensor product on
Rep(SO(3)) satisfy the same axioms as sum and product in a ring. We
phrased this by saying that it was a ring category. This was followed
by the suggestion that module categories over it could be naturally be
defined.

A {\bf module category} structure on a category M over a tensor
category A is given by a functor $ \triangleright : A \times M \rightarrow M$
satisfying the usual consistency condition of a module. 

In \cite{Y1} this definition was augmented by a definition of a module
over a tensor functor.

If $ F:
A\rightarrow B$ is a tensor functor, and M and N are module categories
over A and B respectively, then a functor $ \mu :M \rightarrow N $ is
a module functor over F if it intertwines the action functors on the
two modules.

The definition in \cite{Y1} also includes a coherence diagram (a
pentagon); in the application to the time functor, this will need to
be modified to include the correction factor in the associator, which
we deiscussed above. 

In \cite{E} it was shown that the irreducible module categories over Rep( SU(2)) and
its quantum deformation $Rep(U_q SU(2))$ are classified by simply laced
Dynkin diagrams, i.e. correspond to the ADE series of simple Lie
groups.

In the case of q a root of unity, the rank of the group must correspond
to the level of the root.

This discovery was motivated by a series of discoveries in conformal
field theory \cite{cft} which showed that conformal field theories
built out of combinations of representations of the Kac-Moody algebra
over SU(2) similarly corresponded to the ADE diagrams. The excitations
of the extended CFT's are particular combinations of the fundamental
fields of the basic ones.

The classification of irreducible module categories over Rep(SO(3)) can be
explained by elementary means. If we are given a discrete subgroup $
\Gamma$ of SO(3), then any representation of SO(3) can be considered
as a representation of $ \Gamma$ by restriction. The tensor product in
$ Rep( \Gamma) $ then gives us an action of Rep(SO(3)) on $Rep(
\Gamma) $. Furthermore, it is known that all module categories over
Rep(SO(3)) are so constructed \cite{E}.

The operation of restriction of a representation to a subgroup has an
adjoint called induction. The theory of induced representations allows
us to regard objects in the model categories we are constructing as
objects in the original representation categories of the Lie
groups SO(3) etc (simple objects often get mapped to large direct
sums). 
We want to use this to associate to excitations in the
theory we build out of modules geometric variables in our ``base''
theory, i.e to regard them as having gravitational fields.

The finite subgroups of SO(3) are well known. They consist of two
infinite families, the cyclic and dihedral groups, and three special
groups corresponding to the symmetries of the platonic solids (dual
solids have the same symmetries. As abstract groups, these three are
just $A_4, S_4$ and $A_5$. They are also referred to as T O and I
(tetrahedron, octahedron, and icosahedron; whose symmetry groups they
are) in the context of finite rotation groups. 

(Now the classification of module categories over the representation
category of the quantum group $U_q(SL(2))$, or equivalently of the
corresponding Kac-Moody algebra requires more abstract arguments, but
ends up with the same form, except for a restriction at q a root of
unity \cite{E}. Thus the extension of our program to q-deformed
gravity \cite{Sm} would be straightforward.)

So a module over the time functor would be given by a combination of discrete
subgroups of SO(3) and SO(4) or SO(3,1) depending on the signature we
consider, with a suitable functor connecting their representation
categories. We discuss this below.

We can now see how the group $A_4$ appears naturally as a module category over
Rep(SO(3)), the first step to connecting our module category program
to particle Physics.

Now we would like to explain the ADE classification of the finite
rotation groups, and in particular
see if $A_4$
is somehow related to $E^6$ to unite the picture which
appears from neutrino Physics. This leads us to the Mckay
correspondence in the next section.

\bigskip

{\bf The classical and quantum Mckay correspondence. Applications to
  State Sum Models.}

\bigskip
In \cite{Mac}, Mckay discovered a mysterious relationship between the
finite subgroups of SO(3), or more precisely, their double covers in
SU(2), and the simply laced Lie groups, i. e. ADE in the well known
classification. The representations of the finite groups correspond to
the vertices of the Dynkin diagram of the corresponding Lie group, and
the action of the two dimensional representation of SU(2) restricted
to a representation of the finite group by the tensor product gives
the edges of the diagram. To be more precise, they correspond to the
extended diagram, which is the Dynkin diagram of the corresponding
Kac-Moody algebra. 

In this correspondence, the three non-planar finite rotation groups,
T,O and I correspond to the exceptional algebras $ E_6, E_7, E_8$. So
we see the two pieces of symmetry which appear in the theory of
elementary particles when we include neutrino oscillation Physics correspond to
one another according to McKay!

The term ``Mckay correspondence'' has evolved in the literature to denote a
rich variety of constructions which attempt to shed light on what at
first seems to be an astonishing coincidence. The ADE classification
of the representation categories of the quantum group $U_qSU(2)$ is sometimes
called the ``quantum Mckay correspondence.'' Module categories
can be thought of as a q-version of discrete subgroups.

A natural goal in this field of Mathematics has been to find a
construction which would begin with the relevant finite group and
directly construct the corresponding Kac-Moody algebra.

This has been accomplished in two ways. There is an algebraic
construction due to I. Frenkel and his students \cite{Fr2} and a
geometric one due to Nakajima \cite {Nak}.

Both of these constructions essentially involve construction a bosonic
Fock space from the finite group. The Frenkel construction takes the
sum of the category of representations of the wreath product of the
finite group of all orders. The Nakajima construction begins with the
quotient of a vector space by the action of the group, then considers
symmetric products of all orders of the space with itself
(configuration spaces). Both constructions give us the fundamental
representation of the Kac-Moody algebra.

This suggests that in a state sum model based on a module category
over Rep(So(3)) gauge field symmetries under $E^6$ could appear in the
continuum limit as collective states on large faces composed of many
elementary faces which had quantum states corresponding to the
representations of the finite group T.

We do not yet understand the mathematical details of the construction
of module functors well enough to say how the two kinds of representations would
intertwine. There are still several possibilities for the geometry
corresponding to a module over the time functor. We outline the
current state of our understanding below.
\bigskip

{\bf Towards Picking a Theory}

\bigskip

A module over the time functor would consist of a discrete subgroup G of
SO(3), another discrete subgroup H of SO(3,1) or SL(2,C), and a
functor from Rep G to Rep H which intertwines the two actions and
lifts the time functor. We have not yet completed the mathematical
analysis of this condition, so this work is still partly
programmatic. The details will be explored in a Mathematics paper.

The analogous problem in the Euclidean signature is easier to
tackle. Since SO(4) splits into a left and right copy of SU(2)
(factors of $Z_2$ to one side), a finite subgroup H of SO(4) has two
homomorphisms to finite subgroups of $G_L, G_R$ of SO(3) \cite{book}. 
In the simplest case, H is just the product of the two subgroups, and
if they are isomorphic then sending a representation $ \gamma $ of G
to the representation $ \gamma_L \otimes \gamma_R $ is clearly a module
over $F_t$.

The Lorentzian signature is more complex, and seems to offer more
possibilities. A discrete subgroup of SL(2,C) is called a Kleinian
group, and a rich literature on them exists.

So if G is the symmetry group of the tetrahedron in SO(3), and H is a
Kleinian group, what condition would we expect to allow a module
functor over $F_{ \gamma}$ to connect them? The most obvious choice is
for an injective homomorphism to exist $ G \rightarrow H$.

This turns out to be a mathematically interesting condition. It is
equivalent to H having a ``tetrahedral point'' in hyperbolic space.
In fact, the Kleinian group with the minimal hyperbolic covolume
has such a point \cite{G}. The implications of this choice for
particle Physics still need to be explored.

Another possibility is just to let H=G =T, i. e. to embed the subgroup
T of SO(3) into SO(3,1) directly.

Now there are several ways we can imagine $T \otimes E^6 $ symmetry
arising in these scenarios. One is to use just G=H=T, and have the
$E^6$ symmetry arise in the continuum limit by the quantum Mckay
construction of Frenkel et. al. \cite{Fr2}.

Another possibility is via a more complex Kleinian group. Kleinian
groups of finite covolume correspond to Hyperbolic Kac-Moody algebras
\cite{N}. Some of these look like copies of $E^6$ but with the copy of
SU(2) at a node replaced by a nilpotent Lie algebra. Could this give
rise to a symmetry breaking mechanism? 

A more subtle possibility would be to use the q-deformed gravity
proposal in \cite{Sm}. As we explained, the module categories over
quantum groups are very similar to the ones for the Lie groups,
although they need more abstract treatment \cite{E}. The suggestion in
\cite{C1} to pass to a TQFT in order to include the cosmological
constant as in \cite{Sm} is therefore also possible, although the
categorical quantum analog of Kleinian groups has not been studied
yet.

So the suggestion we have made to construct new grand unified theories
is as yet mostly a program, but a well defined one at least. We have
at least commenced work on the necessary mathematical foundations for
it.

\bigskip

{\bf Summary}

\bigskip

The standard model of particle Physics resembles a jigsaw puzzle. The
small number of symmetries and representations seem to fit together
into a simple whole, but none of the attempts to find the whole have
succeeded.

As usual when a puzzle is not solved, we begin to suspect that some
pieces are missing. The intuition of the author is that discrete
symmetry relating the maddening 3 generations may well be the piece.

If that is correct, tensor categories are a natural setting to unite
Lie group and discrete group symmetries, and module categories are a
nice bridge between them.

If one of the models we construct from Kleinian groups can explain why
quarks and neutrinos come in different representations of SU(3) and T
both, then the approach will become compelling. 

\bigskip

{\bf ACKNOWLEDGEMENT} The author wishes to thank Marc Lachieze-Ray for
his hospitality at Universite Paris 7. The participants in my seminar
at KSU, including Dany Majard and especially David Yetter made many
helpful comments.

\end{document}